\documentclass[singlespacing]{elsart}

\usepackage{graphicx}

\usepackage{amssymb}
\journal{Journal of Molecular Spectroscopy}
\begin{document}

\begin{frontmatter}

\title{Calculation of IR absorption intensities for hydrogen bond from exactly solvable Schr\"odinger equation}

\author{ A.E. Sitnitsky},
\ead{sitnitsky@kibb.knc.ru}

\address{Kazan Institute of Biochemistry and Biophysics, FRC Kazan Scientific Center of RAS, P.O.B. 30,
420111, Russian Federation. e-mail: sitnitsky@kibb.knc.ru}

\begin{abstract}
A theoretical description of IR spectroscopy data for a hydrogen bond (HB) is constructed on the base of trigonometric double-well potential for which an exact analytic solution of the one-dimensional Schr\"odinger equation (SE) is available. The wave functions (full orthogonal basis) are expressed via the spheroidal function while its spectrum of eigenvalues yields the corresponding energy levels (both special functions are implemented in {\sl {Mathematica}}). Then an approximate solution of two-dimensional SE taking into account the excitation state of heavy atoms stretching mode in HB is obtained. It is constructed by decomposing over the above mentioned basis within the framework of standard adiabatic separating the proton motion from that of the heavy atoms. We exemplify the general theory by calculating the IR relative absorption intensities for HB in the Zundel ion ${\rm{H_5O_2^{+}}}$ (oxonium hydrate).
\end{abstract}

\begin{keyword}
double-well potential, spheroidal function, Zundel ion.
\end{keyword}
\end{frontmatter}
\section{Introduction}
The nature and origin of hydrogen bonds (HBs) remain to be a subject of intensive researches (see, e.g., recent reviews \cite{Zha18}, \cite{Lub19}, \cite{Sch19} and refs. therein). In a commonly used approach the proton position in HB is modeled by the Schr\"odinger equation (SE) with a double-well potential (DWP) \cite{Som62}, \cite{Zun69}, \cite{Wie70}, \cite{Jan72}, \cite{Jan73}, \cite{Eck87}, \cite{Kry91}, \cite{Kry92}, \cite{Ven97}, \cite{Zun00}, \cite{McK14}, \cite{Sit17}, \cite{Sit19}. In particular it is applied to the investigation of IR-absorption by HB \cite{Jan72}, \cite{Jan73}, \cite{Eck87}, \cite{Joh98}, \cite{Ath17}. The problem can be regarded in a wider perspective of SE for a quantum particle in DWP as an omnipresent model in physics and chemistry \cite{Jel12}, \cite{Xie12}, \cite{Dow13}, \cite{Che13}, \cite{Har14}, \cite{Ish16}, \cite{Sit171}, \cite{Sit18}, \cite{Don18}, \cite{Don181}, \cite{Don182}, \cite{Don183}, \cite{Don19}, \cite{Don191}. Up to recent time researchers dealing with practical problems involving one-dimensional DWPs had to resort to numerical solution of the corresponding SE at their theoretical interpreting experimental data. The reason was in the lack of a convenient DWP for which SE would have an analytic solution (see \cite{Jel12} and refs. therein). The used DWPs for HB were usually composed of polynomials, exponentials (e.g., the double Morse potential) or their combinations. For all such DWPs the corresponding SE was treated by
quasi-classical (WKB) approximation or numerically with ensuing inconvenience in usage and scanning the parameter space of the model. Besides WKB was shown to yield very inaccurate results in the case of low-barrier DWP (taking place for low-barrier HBs) for which the ground state doublet is close to the barrier top \cite{Sit18}. A notable exception is the double Morse potential belonging to a quasi-exactly solvable type for which exact expressions are derived via the functional Bethe ansatz for a limited number of energy levels \cite{Agb11}. Unfortunately these expressions are very difficult for usage. However the situation in the theory changed drastically during last years with the appearance of a number of DWPs for which analytic solutions were obtained via the confluent Heun's function (CHF) \cite{Xie12}, \cite{Dow13}, \cite{Che13},\cite{Har14}, \cite{Sit17}, \cite{Sit171}, \cite{Don18}, \cite{Don181}, \cite{Don182}, \cite{Don183}, \cite{Don19}, \cite{Don191}, \cite{Don192}, \cite{Don20} and the spheroidal function (SF) \cite{Sit171}, \cite{Sit18}.

Trigonometric DWP \cite{Sit17} for which an analytic solution of SE via SF is available \cite{Sit171}, \cite{Sit18} is beneficially distinguished in the regard of convenience for usage. The history of this potential can be traced back to the monograph on SF \cite{Kom76} where eq. (1.9) along with its explicit solution is presented. This eq. is actually SE with trigonometric DWP although this fact was not recognized and no physical consequences or applications of this form of SE were discussed in \cite{Kom76}. Much later mathematical aspects of SE with various trigonometric potentials were considered in \cite{Sch16}. As well no special attention to trigonometric DWP or its physical implications was paid. However trigonometric DWP is extremely convenient for applications because SF is a well described special function \cite{Kom76} implemented in the mathematical software package {\sl {Mathematica}} along with its spectrum of eigenvalues. As a result the calculation of the energy levels becomes an automatic (at a click) procedure. Earlier trigonometric DWP was applied to an asymmetric hydrogen bond in ${\rm{KHCO_3}}$ \cite{Sit17}, inversion of an ammonia molecule ${\rm{NH_3}}$ \cite{Sit171}, \cite{Cai20}, ring-puckering vibration in 1,3-dioxole and  2,3-dihydrofuran \cite{Sit18} and calculations of the polarizability of HB in chromous acid (CrOOH) and potassium dihydrogen phosphate (${\rm{KH_2PO_4}}$) \cite{Sit19}. The aim of the present article is to show that the above solution enables one to calculate conveniently the IR absorption intensities of HB. In particular we show that trigonometric DWP is suitable for the treatment of available literature data on HB in the Zundel ion ${\rm{H_5O_2^{+}}}$ (oxonium hydrate). The Zundel ion (in which the proton is equally shared between two water molecules ${\rm{H_2O\cdot\cdot\cdot H \cdot\cdot\cdot OH_2}}$) seems to be an excellent object to exemplify the general theory because on the one hand it has been investigated by IR spectroscopy \cite{Zun69}, \cite{Wie70}, \cite{Jan72}, \cite{Zun00}, \cite{Ven99}, \cite{Ven01}, \cite{Kul13} in conjunction with quantum chemical {\it ab initio} calculations \cite{Hua05}, \cite{Yu16}. Also there are results of the density-functional theory \cite{Nat15} that however yield different values for the parameters of DWP compared to those of \cite{Yu16}. On the other hand for the Zundel ion the distance between the oxygen atoms is not a fixed and predetermined value but can be varied in a wide range. The choice of the object enables the capabilities of our approach to come into full light.

The paper is organized as follows. In preliminary Sec.2 we briefly summarize the results of \cite{Sit18} to introduce the designations and make the present article a self-contained one. In Sec. 3 the solution of two-dimensional Schr\"odinger equation is presented and the eigenvalues are found. In Sec. 4 IR absorption intensities are considered. In Sec. 5 the results are discussed and the conclusions are summarized.
\section{Solution of one-dimensional Schr\"odinger equation with trigonometric DWP}
In dimensional units the one-dimensional SE for a quantum particle with the reduced mass $M$ has the form
\begin{equation}
\label{eq1} \frac{d^2 \psi \left(X\right)}{dX^2}+\frac{2M}{\hbar^2}\left[E-V\left(X\right)\right]\psi \left(X\right)=0
\end{equation}
where $-L \leq X \leq L$ and $V\left(X\right)$ is a DWP. The latter is assumed to be infinite at the boundaries of the finite interval for the spatial variable $X=\pm L$.
The dimensionless values for the distance $x$, the potential $U(x)$ and the energy $\epsilon$ are introduced as follows
\begin{equation}
\label{eq2} x=\frac{\pi X}{2L}\ \ \ \ \ \ \ \ \ \ \ \ \ \ \ \ \ U(x)=\frac{8ML^2}{\hbar^2 \pi^2}V\left(X\right)\ \ \ \ \ \ \ \ \ \ \ \ \ \epsilon=\frac{8ML^2E}{\hbar^2 \pi^2}
\end{equation}
where $-\pi/2\leq x \leq \pi/2$. As a result we obtain dimensionless SE
\begin{equation}
\label{eq3}\psi''_{xx} (x)+\left[\epsilon-U(x)\right]\psi (x)=0
\end{equation}
In the symmetric case the trigonometric DWP has the form \cite{Sit18}
\begin{equation}
\label{eq4} U(x)=\left(m^2-\frac{1}{4}\right)\ \tan^2 x-p^2\sin^2 x
\end{equation}
Here $m$ is an integer number and $p$ is a real number. The examples of trigonometric DWP are given in Fig.1, Fig.2$\ $ and Fig.3.
The solution of (\ref{eq3}) with DWP (\ref{eq4}) is \cite{Sit18}
\begin{equation}
\label{eq5} \psi_q (x)=\cos^{1/2} x\ \bar S_{m(q+m)}\left(p;\sin x\right)
\end{equation}
Here $q=0,1,2,...$ and $\bar S_{m(q+m)}\left(p;s\right)$ is the normalized angular prolate SF. The latter is related to $\rm{SpheroidalPS}[(q+m),m,ip,s]$ implemented in {\sl {Mathematica}} as
\[
\bar S_{m(q+m)}\left(p;s\right)=\rm{SpheroidalPS}[(q+m),m,ip,s]\times
\]
\begin{equation}
\label{eq6} \left\{\int_{-1}^{1}ds\ \Biggl(\rm{SpheroidalPS}[(q+m),m,ip,s]\Biggr)^2\right\}^{-1/2}
\end{equation}
The energy levels are determined by the relationship
\begin{equation}
\label{eq7}
\epsilon_q=\lambda_{m(q+m)}\left(p\right)+\frac{1}{2}-m^2-p^2
\end{equation}
Here $\lambda_{m(q+m)}\left(p\right)$ is the spectrum of eigenvalues for $\bar S_{m(q+m)}\left(p;s\right)$. It is implemented in {\sl {Mathematica}} as $\lambda_{m(q+m)}\left(p\right)\equiv \rm{SpheroidalEigenvalue}[(q+m),m,ip]$.
\section{Solution of two-dimensional Schr\"odinger equation}
At IR-absorption transitions in HB the excitation state of heavy atoms vibration (O-O stretching mode in the case of the Zundel ion) can be changed. For this reason one has to use a two-dimensional SE where the vibrational mode with the low frequency $\Omega \sim 10^2\ {\rm cm^{-1}}$ in a harmonic potential for the coordinate $Z$ of heavy atoms in HB is explicitly taken into account \cite{Jan72}, \cite{Jan73}, \cite{Eck87}. Following these articles and also \cite{Ven99}, \cite{Ven01} we construct a model Hamiltonial for the fragment ${\rm{A_1H \cdot \cdot \cdot A_2}}$ treating other degrees of freedom as a thermal bath. Following \cite{May11}, \cite{Gie06} we add to trigonometric DWP $V\left(X\right)$ the interaction of the proton motion with the vibrational mode of the heavy atoms
\[
\Biggl\{\frac{\hbar^2}{2}\left[\frac{1}{\mu }\frac{d^2 }{dZ^2}+\frac{1}{M}\frac{d^2}{dX^2}\right]+E-V\left(X\right)-\frac{\mu \Omega^2}{2}Z^2-
\]
\begin{equation}
\label{eq8} \lambda F\left(X,Z\right)\Biggr \}\Phi \left(X,Z\right)=0
\end{equation}
Here $\lambda$ is a coupling constant, $F\left(X,Z\right)$ is an arbitrary function that can describe, e.g., a symmetric mode coupling ($F\left(X,Z\right)=ZX^2$), an anti-symmetric mode coupling ($F\left(X,Z\right)=ZX$) or a squeezed coupling ($F\left(X,Z\right)=Z^2X^2$) \cite{May11}, \cite{Gie06} and $\mu$ is the reduced mass of the heavy atoms in HB
\begin{equation}
\label{eq9} \mu=\frac{M_1M_2}{M_1+M_2}
\end{equation}
Here $M_i$ ($i=1,2$) is the mass of the fragment $A_i$ that depending upon a chosen model can be conceived as a single atom or a group of atoms. E.g., for HB in the Zundel ion it can be chosen as the mass of the oxygen or that of the water molecule. In accordance with the above choice of the system under consideration ${\rm{O\cdot\cdot\cdot H \cdot\cdot\cdot O}}$ we conceive $A_i$ as single oxygen atoms, i.e., set further $\mu=M_O/2$. This choice coincides with that of \cite{Jan73}. We introduce dimensionless values as in (\ref{eq2})
\begin{equation}
\label{eq10} z=\frac{\pi Z}{2L};\ \ \ \ \ \ \ \ \ \ \ \Lambda=\frac{8ML^2E}{\hbar^2 \pi^2};\ \ \ \ \ \ \ \ \ \ \ \omega=\frac{4\sqrt{2M\mu}L^2\Omega}{\hbar \pi^2}
\end{equation}
\begin{equation}
\label{eq11} \alpha=\frac{8\lambda ML^2}{\hbar^2 \pi^2};\ \ \ \ \ \ \ \ f\left(x,z\right)=F\left(X(x),Z(z)\right);\ \ \ \ \ \ \ \delta=\frac{M}{\mu}
\end{equation}
and further explicitly take into account that both the wave function $\Phi^{k}(x,z)$ and the energy $\Lambda^k_q$ in this case explicitly depend on the quantum number $k$ quantizing the excitation states of the heavy atoms vibration in HB. We obtain the dimensionless form of the two-dimensional SE
\begin{equation}
\label{eq12} \left\{\delta\frac{d^2 }{dz^2}+\frac{d^2}{dx^2}+\Lambda^k_q-U(x)-\frac{\omega^2z^2}{2}-\alpha f\left(x,z\right)\right \}\Phi^{k}(x,z)=0
\end{equation}
where $U(x)$ is given by (\ref{eq4}). Following a standard approach (see, e.g., \cite{Osa00}) we seek the solution of (\ref{eq12}) by decomposing $\Phi^{k}(x,z)$ over the wave functions $\psi_q (x)$ given by (\ref{eq5}). This procedure actually means that we work within the framework of adiabatic separating the proton motion from that of heavy atoms in HB \cite{Jan73}, \cite{Ven97}, \cite{Zun00}, \cite{May11}, \cite{Gie06}. Although we do not explicitly make use of the small parameter $\delta$ ($\delta=1/8$ in the case of the Zundel ion) but by the above mentioned decomposing we retain the essence of the adiabatic approximation, i.e., first solve SE for the proton (find $\psi_q (x)$) supposing the heavy atoms to be fixed ones and then solve SE for the heavy atoms in an effective potential determined among others by the proton wave function. Thus we set
\begin{equation}
\label{eq13} \Phi^{k}(x,z)=\sum_{q=0}^{\infty}\ \varphi^q_k(z) \psi_q (x)
\end{equation}
For the coefficients $\varphi^q_k(z)$ we obtain a closed and self-consistent system of differential equations
\begin{equation}
\label{eq14} \left[\delta\frac{d^2}{dz^2}-\epsilon_q+\Lambda^k_q-\frac{\omega^2 z^2}{2}-\alpha a_{qq}(z)\right]\varphi^q_k(z)=\alpha\sum_{l=0, l \neq q}^{\infty}\ a_{ql}(z)\varphi^l_k(z)
\end{equation}
where $\epsilon_q$ is given by (\ref{eq7}) and the coefficients $a_{ql}$ are
\begin{equation}
\label{eq15} a_{ql}(z)=\int_{-\pi/2}^{\pi/2}dx\ \psi_q (x)f\left(x,z\right)\psi_l (x)
\end{equation}
We further distinguish the types of mode coupling by the subscript $|^{\{t\}}$ where this sign means $\{s\}$ for the symmetric case, $\{as\}$ for the anti-symmetric one or $\{sq\}$ for the squeezed coupling if it is not indicated explicitly. In our opinion for all of them it is reasonable to go beyond the linear approximation as it will be done for the dipole moment (see discussion before (\ref{eq26})), i.e., to make the replacing $x\longrightarrow \sin x$ at $-\pi/2 \leq x \leq \pi/2$. Thus we further take $f\left(x,z\right)|^{\{s\}}\propto \sin^2 x$, $f\left(x,z\right)|^{\{as\}}\propto \sin x$ and $f\left(x,z\right)|^{\{sq\}}\propto \sin^2 x$. We obtain
\begin{equation}
\label{eq16} \alpha a^{\{s\}}_{qq}(z)=c^{\{s\}}_qz;\ \ \ \ \ \ c^{\{s\}}_q=\frac{2^6\lambda^{\{s\}} ML^5}{\hbar^2 \pi^5}\int_{-1}^{1}d\eta\ \eta^2\ \left[\bar S_{m(q+m)}\left(p;\eta\right)\right]^2
\end{equation}
\begin{equation}
\label{eq17}  \alpha a^{\{as\}}_{qq}(z)=c^{\{as\}}_qz;\ c^{\{as\}}_q=\frac{2^5\lambda^{\{as\}} ML^4}{\hbar^2 \pi^4} \int_{-1}^{1}d\eta\ \eta\ \left[\bar S_{m(q+m)}\left(p;\eta\right)\right]^2= 0
\end{equation}
\begin{equation}
\label{eq18}  \alpha a^{\{sq\}}_{qq}(z)=c^{\{sq\}}_qz^2;\ \ \ c^{\{sq\}}_q=\frac{2^7\lambda^{\{sq\}} ML^6}{\hbar^2 \pi^6}\int_{-1}^{1}d\eta\ \eta^2\ \left[\bar S_{m(q+m)}\left(p;\eta\right)\right]^2
\end{equation}
The main problem in this case is to solve the system (\ref{eq14}) for the functions $\varphi^q_k(z)$ and to find in particular the spectrum $\Lambda^k_q$. We conceive mixing $\varphi^q_k(z)$ with other $\varphi^l_k(z)$ ($l \neq q$) terms (the right-hand sides in (\ref{eq14})) as a small perturbation and in the zero-order approximation neglect it. It will be argued below that in practical cases this approximation is very accurate. Thus for all types of coupling we have some sort of a modified harmonic oscillator equation. We obtain for the symmetric and anti-symmetric mode couplings
\[
\varphi^q_k(z)|^{\{t\}}\approx \exp\left[-\frac{\omega}{2\sqrt{2\delta}}\left(z+\frac{c^{\{t\}}_q}{\omega^2}\right)^2\right]\frac{k!(-2)^k}{(2k)!}\times
\]
\begin{equation}
\label{eq19}
 H_{2k}\left(\left(\frac{2}{\delta}\right)^{1/4}\sqrt {\omega}\left(z+\frac{c^{\{t\}}_q}{\omega^2}\right)\right)
\end{equation}
\begin{equation}
\label{eq20} \Lambda^k_q|^{\{t\}}\approx\epsilon_q-\frac{\left(c^{\{t\}}_q\right)^2}{2\omega^2}+(4k+1)\omega \sqrt{\frac{\delta}{2}}
\end{equation}
where $k=0,1,2,...\ $. For the squeezed coupling we obtain
\[
\varphi^q_k(z)|^{\{sq\}}\approx \exp\left[-\frac{1}{2}z^2\sqrt{\frac{\omega^2+2c^{\{sq\}}_q}{2\delta}}\ \right]\frac{k!(-2)^k}{(2k)!}\times
\]
\begin{equation}
\label{eq21}
 H_{2k}\left(\sqrt {2}z\left(\frac{\omega^2+2c^{\{sq\}}_q}{2\delta}\right)^{1/4}\right)
\end{equation}
\begin{equation}
\label{eq22} \Lambda^k_q|^{\{sq\}}\approx\epsilon_q+(4k+1)\sqrt{\frac{\delta\left(\omega^2+2c^{\{sq\}}_q\right)}{2}}
\end{equation}
Further we conceive the functions $\varphi^q_k(z)|^{\{t\}}$ as being normalized ones
\begin{equation}
\label{eq23} \bar\varphi^q_k(z)|^{\{t\}}=\varphi^q_k(z)|^{\{t\}}\left\{\int_{-\infty}^{\infty}dz\ \Biggl(\varphi^q_k(z)|^{\{t\}}\Biggr)^2\right\}^{-1/2}
\end{equation}
\section{Relative IR absorption intensities for hydrogen bond}
Modern researches on HB in the Zundel ion deal with temperatures up to  $T= 1\ K$ \cite{Kal09}. To extend the theory into low temperature region the so-called harmonic quantum correction factor \cite{Kal09} is introduced that actually replaces the Boltzmann distribution function for the population of the energy levels by the Bose-Einstein one. The relative absorption intensity $I_{ji}$ from the state $j$ with the energy $E_j$ to the state $i$  with the energy $E_i$ \cite{Eck87} taking into account the above correction factor is
\[
 I_{ji}=\frac{E_i-E_j}{\hbar}\mid \mu_{ji}\mid^2\exp\left(-\frac{E_j}{k_BT}\right)\left[1-\exp\left(-\frac{E_j}{k_BT}\right)\right]^{-1}\times
\]
\begin{equation}
\label{eq24}\left[\sum_{n=0}^{\infty}\ \exp\left(-\frac{E_n}{k_BT}\right)\left[1-\exp\left(-\frac{E_n}{k_BT}\right)\right]^{-1}\right]^{-1}
\end{equation}
where $\hbar$ is the reduced Planck constant, $k_B$ is the Boltzmann constant, $T$ is the temperature and $\mu_{ji}$ is the transition dipole matrix element
\begin{equation}
\label{eq25} \mu_{ji}=\int dX\ \Psi^{\ast}_j (X)\ \mu(X)\ \Psi_i (X)
\end{equation}
Here $X$ denotes the quantum particle coordinate (e.g., that of the proton along the O-O axis in O-H $\cdot \cdot \cdot$ O), $\mu(X)$ is the dipole moment and $\Psi_i (X)$ is the wave function of the $i$-th state. The dipole moment $\mu(X)$ is usually assumed to be $\mu(X)=e X=e2Lx/\pi$ in the linear approximation where $e$ is the quantum particle charge.  However the linear approximation for the dipole moment can be valid within the interval of a sufficiently small $x$ only. It is commonly accepted that in reality the dipole moment deflects to slower growth than the linear one (see, e.g., Fig. 10.54 in \cite{Atk09}). The necessity to go beyond the framework of the linear approximation for HB in the Zundel ion was stressed in \cite{Ven01}. We achieve this goal and model such deflection by replacing the linear term by the trigonometric one $x\longrightarrow \sin x$ at $-\pi/2 \leq x \leq \pi/2$. Thus the dipole moment from (\ref{eq25}) with taking into account the above replacement is
\begin{equation}
\label{eq26} \mu(X)=\frac{e2L}{\pi} \sin x
\end{equation}
If the heavy atoms stretching mode is taken into account then the transition dipole matrix element is $\mu^{kn}_{ji}$. Thus the wave function $\Psi_i (X)$ should explicitly include the quantum numbers $k$ and $n$ of the heavy atoms excitation state and also depend on their spatial variable \cite{Jan72}, \cite{Jan73}, \cite{Eck87}, \cite{Ven01}, i.e.,  $Z=2Lz/\pi$ in our notation. In this case we identify the function $\Psi_i (X=2Lx/\pi)$ with the product $\bar\varphi^j_k(z)|^{\{t\}} \psi_j (x)$ so that the transition dipole matrix element (depending on the assumed type of mode coupling $|^{\{t\}}$) is
\begin{equation}
\label{eq27} \mu^{kn}_{ji}|^{\{t\}}\approx\frac{e2L}{\pi}\int_{-\infty}^{\infty}dz\int_{-\pi/2}^{\pi/2}dx\ \bar\varphi^j_k(z)|^{\{t\}} \psi_j (x)\ \sin x\ \bar\varphi^i_n(z)|^{\{t\}} \psi_i (x)
\end{equation}
We denote the dimensionless inverse temperature $\beta$ and the dimensionless relative adsorption intensity $\Gamma^{kn}_{ji}\left(\beta\right)$
\begin{equation}
\label{eq28}
\beta=\frac{\hbar^2\pi^2}{8ML^2k_BT}\ \ \ \ \ \ \ \ \ \ \ \ \ \ \ \ \ \ \ \ \ \ \ \ \ \ \ \ \Gamma^{kn}_{ji}\left(\beta\right)=\frac{2 M}{\hbar e^2}I_{ji}
\end{equation}
The dimensionless relative absorption intensity is
\[
\Gamma^{kn}_{ji}\left(\beta\right)|^{\{t\}}\approx \left[\int_{-\infty}^{\infty}dz\ \bar\varphi^j_k(z)|^{\{t\}} \bar\varphi^i_n(z)|^{\{t\}} \right]^2 \left[\sum_{q=0}^{\infty}\sum_{k=0}^{\infty}\ \frac{\exp \left(-\beta \Lambda^k_q|^{\{t\}}\right)}{1-\exp \left(-\beta \Lambda^k_q|^{\{t\}}\right)}\right]^{-1}\times
\]
\begin{equation}
\label{eq29} \frac{\exp \left(-\beta \Lambda^k_q|^{\{t\}}\right)\left(\Lambda^n_i|^{\{t\}}-\Lambda^k_j|^{\{t\}}\right)}{1-\exp \left(-\beta \Lambda^k_q|^{\{t\}}\right)}\left[\int_{-1}^{1}d\eta\ \eta\ \bar S_{m(j+m)}\left(p;\eta\right)\bar S_{m(i+m)}\left(p;\eta\right)\right]^2
\end{equation}

\section{Results and discussion}
\begin{figure}
\begin{center}
\includegraphics* [width=\textwidth] {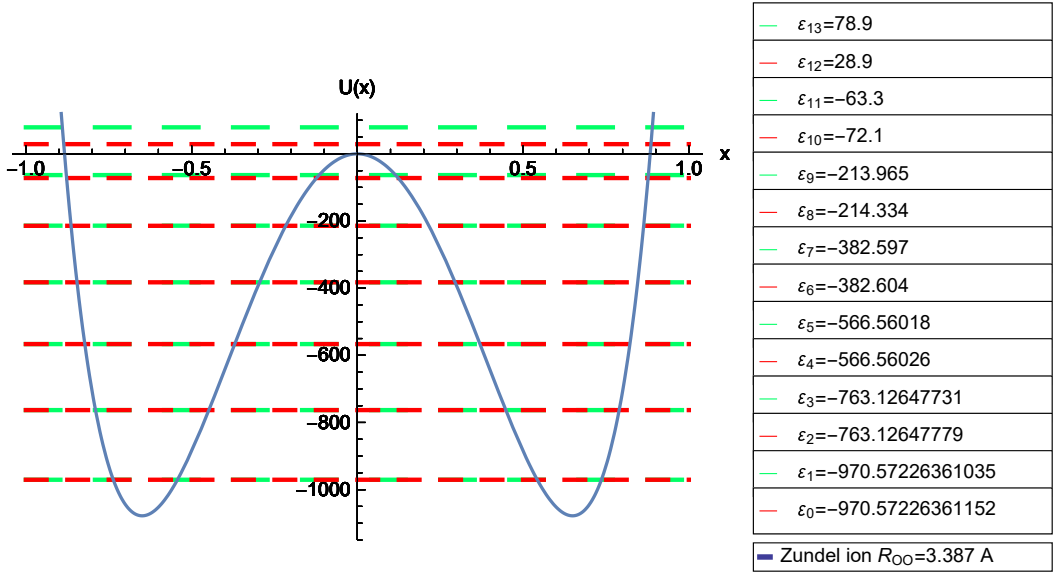}
\end{center}
\caption{The trigonometric double-well potential (\ref{eq4}) at the values of the parameters $m=57$ and $p=89.8295$. The parameters are chosen to describe the potential and the energy levels for the hydrogen bond in the Zundel ion ${\rm{H_5O_2^{+}}}$ (oxonium hydrate) for $R_{OO}=3.387\ \AA$
(experimental data are taken from \cite{Yu16}). The barrier height $B=-U\left(x_{min}\right)=1076.46$ corresponds to $13850\ {\rm cm^{-1}}$ in dimensional units. The splitting of the ground state $\epsilon_1-\epsilon_0=1.17507\cdot 10^{-9}$ corresponds to $1.51187\cdot 10^{-8}\ {\rm cm^{-1}}$ in dimensional units.} \label{Fig.1}
\end{figure}
Fig.1 shows that trigonometric DWP fits the results of quantum chemical calculations for HB in the Zundel ion with $R_{OO}=3.387\ \AA$ from Fig.1 of \cite{Yu16}. From there we obtain that for $R_{OO}=3.387\ \AA$ ($L=3.387/2\ \AA$) the dimensional distance between the minima of DWP is $X_{min}^{(1)}-X_{min}^{(2)}\approx 1.4\ \AA$. The dimensional barrier height is $V\left(X_{max}\right)-V\left(X_{min}\right)\approx 13850\ {\rm cm^{-1}}$. Taking into account that for a proton $M=1\ {\rm amu}$ we obtain with the help of (\ref{eq2}) the dimensionless values for the barrier height $B=-U\left(x_{min}\right)\approx 1076.46$ and width $D=x_{min}^{(1)}-x_{min}^{(2)}\approx 1.2979$. The transformation formulas for the parameters of trigonometric DWP $\{m,p\}$ into $\{B,D\}$ are \cite{Sit19}
\[
p=\frac{\sqrt {B}}{1-\left[\cos\left(D/2\right)\right]^2} \ \ \ \ \ \ \ \ \ \ \ \ \ \ \ \ \ \ \ \ \ \ \ m^2-\frac{1}{4}=\frac{B\left[\cos\left(D/2\right)\right]^4}{\left\{1-\left[\cos\left(D/2\right)\right]^2\right\}^2}
\]
We have $p\approx 89.8295$ and $m\approx 57$. Here it is pertinent to stress the following methodical trick. The implementation of the eigenvalue for SF\\ $\rm{SpheroidalEigenvalue}[(q+m),m,ip]$ in {\sl {Mathematica}} becomes extremely capricious at high values of the parameters $m$ and $p$. One has to take at $m=57$ the value $p = 89.8295000000000000$ instead of $p = 89.8295$ for the software package to yield the ground state doublet. The calculation results in 6 doublets below the barrier top.
\begin{figure}
\begin{center}
\includegraphics* [width=\textwidth] {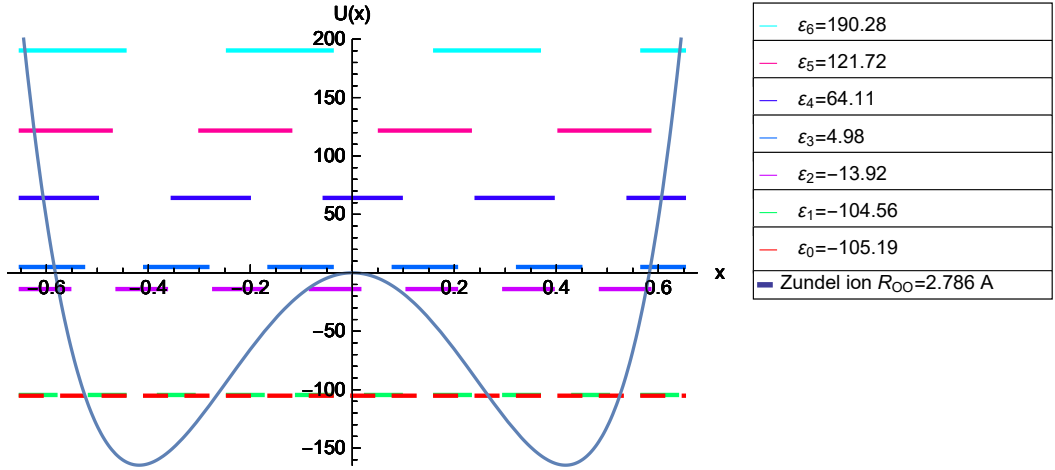}
\end{center}
\caption{The trigonometric double-well potential (\ref{eq4}) at the values of the parameters $m=65$ and $p=77.83$. The parameters are chosen to describe the potential and the energy levels for the hydrogen bond in the Zundel ion ${\rm{H_5O_2^{+}}}$ (oxonium hydrate) for $R_{OO}=2.786\ \AA$
(experimental data are taken from \cite{Yu16}). The barrier height $B=-U\left(x_{min}\right)=163.02$ corresponds to $3100\ {\rm cm^{-1}}$ in dimensional units. The splitting of the ground state $\epsilon_1-\epsilon_0=0.63$ corresponds to $11.98\ {\rm cm^{-1}}$ in dimensional units.} \label{Fig.2}
\end{figure}

In Fig.2 trigonometric DWP fits the results of quantum chemical calculations for HB in the Zundel ion with $R_{OO}=2.786\ \AA$ ($L=2.786/2\ \AA$) presented in Fig.1 of \cite{Yu16}. From there we obtain that the dimensional distance between the minima of DWP is $X_{min}^{(1)}-X_{min}^{(2)}\approx 0.74\ \AA$. The dimensional barrier height is $V\left(X_{max}\right)-V\left(X_{min}\right)\approx 3100\ {\rm cm^{-1}}$. From these values we obtain the dimensionless values for the barrier height $B=-U\left(x_{min}\right)\approx 163.02$ and width $D=x_{min}^{(1)}-x_{min}^{(2)}\approx 0.83$. From here we have $p\approx 77.83$ and $m\approx 65$. The calculation yields the ground state doublet well below the barrier top and a doublet in its vicinity.

In Fig.3 a low-barrier case is presented that fits the results of quantum chemical calculations for HB in the Zundel ion with $R_{OO}=2.536\ \AA$ ($L=2.536/2\ \AA$) from Fig.1 of \cite{Yu16}. From there we obtain that the dimensional distance between the minima of DWP is $X_{min}^{(1)}-X_{min}^{(2)}\approx 0.4\ \AA$. The dimensional barrier height is $V\left(X_{max}\right)-V\left(X_{min}\right)\approx 400\ {\rm cm^{-1}}$. From these values we obtain the dimensionless values for the barrier height $B=-U\left(x_{min}\right)\approx 15.7$ and width $D=x_{min}^{(1)}-x_{min}^{(2)}\approx 0.5$. From here we have $p\approx 65.92$ and $m\approx 62$.
\begin{figure}
\begin{center}
\includegraphics* [width=\textwidth] {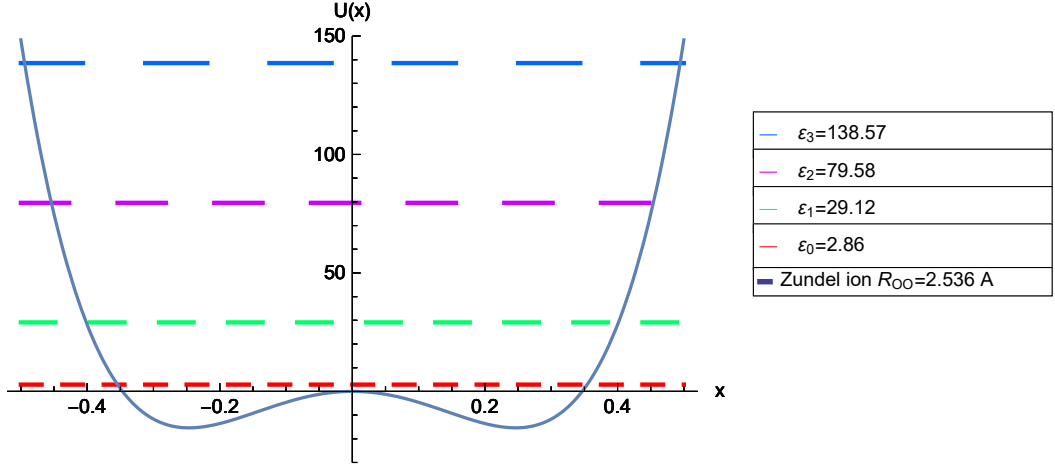}
\end{center}
\caption{The trigonometric double-well potential (\ref{eq4}) at the values of the parameters $m=62$ and $p=65.92$. The parameters are chosen to describe the potential and the energy levels for the hydrogen bond in the Zundel ion ${\rm{H_5O_2^{+}}}$ (oxonium hydrate) for $R_{OO}=2.536\ \AA$
(experimental data are taken from \cite{Yu16}). The barrier height $B=-U\left(x_{min}\right)=15.3$ corresponds to $400\ {\rm cm^{-1}}$ in dimensional units.} \label{Fig.3}
\end{figure}

Unfortunately for the cases $R_{OO}=3.387\ \AA$ and $R_{OO}=2.786\ \AA$ treated quantum chemically in \cite{Yu16} there are no corresponding data on the relative absorption intensities. In contrast the case $R_{OO}=2.536\ \AA$ considered in \cite{Yu16} can be compared with the case $R_{OO}=2.5\ \AA$ for which the pertinent data are calculated in \cite{Jan73}. To do it we normalize the data presented in Table 2 of \cite{Jan73} by the value of the relative absorption intensity for the transition $I_{00\rightarrow 10}$ at T=200 K for $R_{OO}=2.5\ \AA$ while normalize our data by the value $\Gamma^{00}_{01}(\beta=0.069)$ at $\beta=0.069$ (T=200 K). We choose the value of the frequency of O-O stretching mode $\omega=1.4$ corresponding to $\Omega\approx 100\ {\rm cm^{-1}}$. We adopt the case of the symmetric mode coupling as in \cite{Jan73} and carry out calculations in the zero-order approximation. It appears to be very accurate because for $l\neq q$
\[
\frac{\alpha a^{\{s\}}_{ql}(z)}{\alpha a^{\{s\}}_{qq}(z)}=\frac{1}{c^{\{s\}}_q}\frac{2^6\lambda^{\{s\}} ML^5}{\hbar^2 \pi^5}\int_{-1}^{1}d\eta\ \eta^2\ \bar S_{m(q+m)}\left(p;\eta\right)\bar S_{m(l+m)}\left(p;\eta\right)\sim 10^{-19}
\]
Thus the terms in the right-hand side of (\ref{eq14}) can be safely discarded.

Fig.4 for HB in the Zundel ion with $R_{OO}=2.536\ \AA$ shows that the output information of quantum chemical calculations (values of the barrier height and width for potential energy surface) can be conveniently related to IR-spectroscopy data with the help of our approach. The temperature dependence of the relative absorption intensity $\Gamma^{00}_{01}\left(\beta\right)$ with no change in the excitation state of the heavy atoms vibration in HB is compared with analogous data from \cite{Jan73} for the case $R_{OO}=2.5\ \AA$. In the notations of \cite{Jan73} it is $00\rightarrow 10$ where the first number corresponds to state of the proton and the second one corresponds to that of the heavy atoms. We obtain $\Gamma^{00}_{02}\left(\beta\right)=0$ (because of the parity of $\psi_0 (x)$ and $\psi_2 (x)$ at the calculation of the transition dipole matrix element (\ref{eq27})) that agrees with data from \cite{Jan73}. Also in Fig.4 analogous results for some transitions with the change of the heavy atoms vibrational quantum number are presented. As a whole our results are in agreement with the corresponding data from \cite{Jan73}.
\begin{figure}
\begin{center}
\includegraphics* [width=\textwidth] {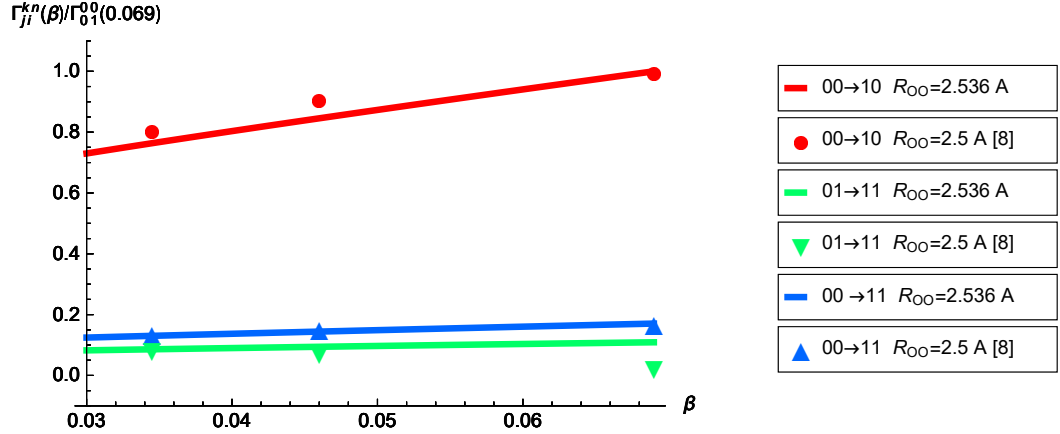}
\end{center}
\caption{Temperature dependence of normalized relative absorption intensities (\ref{eq29}) at the values of the parameters $m=62$ and $p=65.92$ in the case of the symmetric mode coupling (\ref{eq16}). The parameters are chosen to describe the potential and the energy levels for the hydrogen bond in the Zundel ion ${\rm{H_5O_2^{+}}}$ (oxonium hydrate) for $R_{OO}=2.536\ \AA$
(experimental data are taken from \cite{Yu16}). The frequency of O-O stretching mode is $\omega=1.4$ corresponding to $\Omega\approx 100\ {\rm cm^{-1}}$ and the coupling constant is $c^{\{s\}}_0=0.6$ ($\lambda^{\{s\}}\approx 0.16$ au). Also the data for $T=400 K$ ($\beta=0.0345$), $T=300 K$ ($\beta=0.046$) and $T=200 K$ ($\beta=0.069$) taken from \cite{Jan73} for $R_{OO}=2.5\ \AA$ are presented.} \label{Fig.4}
\end{figure}

We conclude that the suggested approach enables one to relate in an analytic form the barrier height and width obtained from quantum chemical {\it ab initio} calculations of the potential energy surface for a hydrogen bond with IR transition frequencies and relative absorption intensities. Our theory on the base of the trigonometric double-well potential does not require numerical solution of the corresponding Schr\"odinger equation or making use of the quasi-classical approximation. Thus it seems to be a useful tool for analyzing IR spectroscopy data that provides considerable simplification of the calculations compared with previously used methods. The validity of our approach is proved by its ability to reproduce available literature data for the hydrogen bond in the Zundel ion.

Acknowledgements. The author is grateful to Prof. Yu.F. Zuev for helpful discussions. The work was supported from the government assignment for FRC Kazan Scientific Center of RAS.


\newpage
\begin{thebibliography}{00}
\bibitem{Zha18}
Z. Zhang, D. Li, W. Jiang, Z. Wang, The electron density delocalization of hydrogen bond
systems, Advances in Physics: X, 3 (2018) 1428915.
\bibitem{Lub19}
S.C.C. van der Lubbe, C.F. Guerra, The nature of hydrogen bonds: A delineation of the role of
different energy components on hydrogen bond strengths and
lengths, Chem.Asian J. 14 (2019) 2760-2769.
\bibitem{Sch19}
S. Scheiner, Forty years of progress in the study of the hydrogen bond, Struct.Chem.  30 (2019) 1119-1128.
\bibitem{Som62}
R.L. Somorjai, D.F. Hornig, Double minimum potentials in hydrogen bonded solids, J.Chem.Phys. 36 (1962) 1980-1987.
\bibitem{Zun69}
G. Zundel, Hydration and intermolecular interaction, Academic Press, 1969.
\bibitem{Wie70}
E.G. Weidemann and G. Zundel, Field-dependent mechanism of anomalous proton conductivity
and the polarizability of hydrogen bonds with tunneling protons, Z. Naturforsch., Teil A, 25 (1970) 627-634.
\bibitem{Jan72}
R. Janoschek, E.G. Weidemann, H. Pfeiffer, G. Zundel, Extremely high polarizability of hydrogen bonds, J.Amer.Chem.Soc. 94 (1972) 2387-2396.
\bibitem{Jan73}
R. Janoschek, E.G. Weidemann, G. Zundel, Calculated frequencies and intensities associated with coupling of the proton motion with the hydrogen bond stretching vibration in a double minimum potential surface, J.Chem.Soc., Faraday Transactions 2: Mol.Chem.Phys. 69 (1973) 505-520.
\bibitem{Eck87}
M. Eckert, G. Zundel, Proton polarizabity, dipole moment, and proton transitions of an ${\rm{AH \cdot \cdot \cdot B \rightleftharpoons A^{-} \cdot \cdot \cdot H^{+}B}}$
proton-transfer hydrogen bond as a function of an external electrical field: an ab
initio SCF treatment, J.Phys.Chem. 91 (1987) 5170-5177.
\bibitem{Kry91}
E.S. Kryachko, M. Eckert, G. Zundel, Study of tunnelling in symmetrical double-Morse
hydrogen bonds via the instanton-soliton approach: large polarizability and isotopic effect, J.Mol.Struct. (Theochem), 235 (1991) 157-183.
\bibitem{Kry92}
E.S. Kryachko, M. Eckert, G. Zundel, An approach, still analytical, to the study of proton
tunneling in symmetrical hydrogen bonds, J.Mol.Struct. 270 (1992) 33-65.
\bibitem{Ven97}
M.V. Vener, N.D. Sokolov, On the adiabatic separation of the vibrational variables
of a hydrogen-bonded ${\rm{AHA}}$ fragment with a symmetric double-well potential, Chem.Phys.Lett. 264 (1997) 429-434.
\bibitem{Zun00}
G. Zundel, Hydrogen bonds with large proton polarizability and proton transfer processes in electrochemistry and biology, Adv. Chem.Phys.
111 (2000) 1-217.
\bibitem{McK14}
R.H. McKenzie, C. Bekker, B. Athokpam, S.G. Ramesh, Effect of quantum nuclear motion on hydrogen bonding, J.Chem.Phys. 140 (2014) 174508.
\bibitem{Sit17}
A.E. Sitnitsky, Exactly solvable Schr\"odinger equation with double-well potential for hydrogen
bond, Chemical Physics Letters 676C (2017) 169-173.
\bibitem{Sit19}
A.E. Sitnitsky, Analytic treatment of IR-spectroscopy data for double well potential, Computational and Theoretical Chemistry 1160 (2019) 19-23.
\bibitem{Joh98}
P.G. Johannsen, Vibrational states and optical transitions in hydrogen bonds, J.Phys.:Condens.Matter 10 (1998) 2241-2260.
\bibitem{Ath17}
B. Athokpam, S.G. Ramesh, R.H. McKenzie, Effect of hydrogen bonding on the infrared absorption intensity of OH
stretch vibrations, Chem.Phys. 488-489 (2017) 43-54.
\bibitem{Jel12}
V. Jelic, F. Marsiglio, The double-well potential in quantum mechanics: a simple, numerically exact formulation, Eur. J. Phys. 33 (2012) 1651-1666.
\bibitem{Xie12}
Q.-T. Xie, New quasi-exactly solvable double-well potentials, J.Phys. A: Math. Theor. 45 (2012) 175302.
\bibitem{Dow13}
C.A. Downing, On a solution of the Schr\"odinger equation with a hyperbolic double-well potential, J.Math.Phys. 54 (2013) 072101.
\bibitem{Che13}
B.-H. Chen, Y. Wu, Q.-T. Xie, Heun functions and quasi-exactly solvable double-well
potentials, J.Phys. A: Math. Theor. 46 (2013) 035301.
\bibitem{Har14}
R.R. Hartmann, Bound states in a hyperbolic asymmetric double-well, J.Math.Phys. 55 (2014) 012105.
\bibitem{Ish16}
A. Ishkhanyan, Schr\"odinger potentials solvable in terms of the confluent Heun functions, Theor.Math.Phys. 188 (2016) 980-993.
\bibitem{Sit171}
A.E. Sitnitsky, Analytic description of inversion vibrational mode for ammonia molecule, Vibrational Spectroscopy 93 (2017) 36-41.
\bibitem{Cai20}
Caio M. Porto, Nelson H. Morgon, Analytical approach for the tunneling process in double well potentials using
IRC calculations,  Computational and Theoretical Chemistry 1187 (2020) 112917.
\bibitem{Sit18}
A.E. Sitnitsky, Analytic calculation of ground state splitting in symmetric double well potential, Computational and Theoretical Chemistry 1138 (2018) 15-22.
\bibitem{Don18}
Q. Dong, F.A. Serrano, G.-H. Sun, J. Jing, S.-H. Dong, Semiexact solutions of the Razavy potential, Adv. High Energy Phys.
(2018) 9105825.
\bibitem{Don181}
S. Dong, Q. Dong, G.-H. Sun, S. Femmam, S.-H. Dong, Exact solutions of the Razavy cosine type potential, Adv. High Energy Phys.
(2018) 5824271.
\bibitem{Don182}
Q. Dong, G.-H. Sun, J. Jing, S.-H. Dong, New findings for two new type sine hyperbolic potentials, Phys.Lett. A383 (2019) 270-275.
\bibitem{Don183}
Q. Dong, S.-S. Dong, E. Hern\'andez-M\'arquez, R. Silva-Ortigoza, G.-H. Sun, S.-H. Dong, Semi-exact solutions of Konwent potential, Commun.Theor.Phys. 71 (2019) 231-236.
\bibitem{Don19}
Q. Dong, A.J. Torres-Arenas, G.-H. Sun, Camacho-Nieto, S. Femmam, S.-H. Dong, Exact solutions of the sine hyperbolic type potential, J.Math.Chem. 57 (2019) 1924-1931.
\bibitem{Don191}
Q. Dong, G.-H. Sun, M. Avila Aoki, C.-Y. Chen, S.-H. Dong, Exact solutions of a quartic potential, Mod.Phys.Lett. A 34 (2019) 1950208.
\bibitem{Don192}
Qian Dong, Guo-Hua Sun, N. Saad, Shi-Hai Dong, Exact solutions of a nonpolynomial oscillator related to isotonic
oscillator, Eur.Phys.J. Plus 134 (2019) 562-569.
\bibitem{Don20}
Guo-Hua Sun, Chang-Yuan Chen, Hind Taud, C.Y\'a\~nez-M\'arquez, Shi-Hai Dong, Exact solutions of the 1D Schr\"odinger equation with the Mathieu potential, Phys.Lett. A384 (2020) 126480.
\bibitem{Agb11}
D. Agboola, Quasi-exactly solvable double Morse potential and
proton tunnelling in hydrogen bonded crystals, presentation (unpublished results).
\bibitem{Kom76}
I.V. Komarov, L.I. Ponomarev, S.Yu. Slavaynov, Spheroidal and Coloumb spheroidal functions, Moscow, Science, 1976.
\bibitem{Sch16}
A. Schulze-Halberg, Trigonometric potentials arising from the spheroidal equation: supersymmetric partners and integral formulas, Eur. Phys. J. Plus 131 (2016) 202.
\bibitem{Ven99}
M.V. Vener, J. Sauer, Quantum anharmonic frequencies of the ${\rm{O \cdot \cdot \cdot H \cdot \cdot \cdot O}}$
fragment of the ${\rm{H_5O_2^{+}}}$ ion: a model three-dimensional study, Chem.Phys.Lett. 312 (1999) 591-597.
\bibitem{Ven01}
M.V. Vener, O. K\"uhn, J. Sauer, The infrared spectrum of the ${\rm{O \cdot \cdot \cdot H \cdot \cdot \cdot O}}$ fragment of ${\rm{H_5O_2^{+}}}$: Ab initio classical
molecular dynamics and quantum 4D model calculations, J.Chem.Phys. 114 (2001) 240-249.
\bibitem{Kul13}
W. Kulig, N. Agmon, A 'clusters-in-liquid' method for calculating infrared spectra identifies the proton-transfer mode in acidic aqueous solutions,
Nature Chemistry 5 (2013) 29-35.
\bibitem{Hua05}
X. Huang, B.J. Braams, J.M. Bowman, Ab initio potential energy and dipole moment surfaces for ${\rm{H_5O_2^{+}}}$, J.Chem.Phys. 122 (2005) 044308.
\bibitem{Yu16}
Q. Yu, J.M. Bowman, How the Zundel ${\rm{H_5O_2^{+}}}$ potential can be used to predict the proton
stretch and bend frequencies of larger protonated water clusters, J.Phys.Chem.Lett. 7 (2016) 5259-5265.
\bibitem{Nat15}
S.K. Natarajan, T.Morawietz, J. Behler, Representing the potential-energy surface of
protonated water clusters by high-dimensional neural network potentials, Phys.Chem.Chem.Phys. 17 (2015) 8356.
\bibitem{Osa00}
I.S. Osad'ko, Selective spectroscopy of single molecules, FIZMATLIT, Moscow, 2000.
\bibitem{May11}
V. May, O. K\"uhn, Charge and energy transfer dynamics in molecular systems, 3-d ed., Wiley, 2011.
\bibitem{Gie06}
K. Giese, M. Petkovi\'c, H. Naundorf, O. K\"uhn, Multidimensional quantum dynamics and infrared spectroscopy of
hydrogen bonds, Physics Reports 430 (2006) 211-276.
\bibitem{Kal09}
M. Kaledin, A.L. Kaledin, J.M. Bowman, J. Ding, K.D. Jordan, Calculation of the vibrational spectra of ${\rm{H_5O_2^{+}}}$ and its deuterium-substituted
isotopologues by molecular dynamics simulations, J.Phys.Chem. A 113 (2009) 7671-7677.
\bibitem{Atk09}
P. Atkins, J. de Paula, R.Friedman,  Quanta, Matter, and Change. A molecular approach to physical chemistry, Freeman, 2009.
\end{thebibliography}
\end{document}